\documentclass[12pt]{revtex4}
\usepackage{epsfig,amsbsy,bm,amssymb}
\usepackage{hyperref}
\usepackage{graphicx}
\usepackage{float}
\usepackage{amsmath}

\newcommand{\isum}%
{\mathop{\hbox{$\displaystyle\sum\kern-13.2pt\int\kern1.5pt$}}}
\renewcommand{\r}{{\bm r}}

\newcommand{\p}{{\bm p}}

  \newcommand{\A}{{\bm A}}
	\newcommand{\y}{{\bm y}}
	
   \newcommand{\n}{{\bm n}}
  \newcommand{\e}{{\bm e}}
  \newcommand{\ve}{{\bm v}}

\newcommand{\bt}{\begin{tabular}}
\newcommand{\et}{\end{tabular}}

\newcommand{\eref}[1] {(\ref{#1})}
\newcommand{\Eref}[1] {Eq.~(\ref{#1})}

\newcommand{\Fref}[1] {Fig. \ref{#1}}

\newcommand{\br}{\begin{eqnarray*}}
\newcommand{\er}{\end{eqnarray*}}
\newcommand{\ba}{\begin{eqnarray}}
\newcommand{\ea}{\end{eqnarray}}
\newcommand{\be}{\begin{equation}}
\newcommand{\ee}{\end{equation}}
\newcommand{\hs}{\hspace*}

\newcommand{\rs}{\resizebox}
\newcommand{\bp}{\begin{minipage}}
\newcommand{\ep}{\end{minipage}}

\begin{document}
\thispagestyle{empty}

\bibliographystyle{apsrev}



\title{Relativistic calculation of non-dipole effects in high harmonic generation}

\date{\today}

\author{I. A. Ivanov$^{1}$}
\email{igorivanov@ibs.re.kr}

\author{Kyung Taec Kim$^{1,2}$}
\email{kyungtaec@gist.ac.kr}

\affiliation{$^1$ Center for Relativistic Laser Science, Institute for
  Basic\ Science (IBS), Gwangju 61005, Republic of Korea}
	
\affiliation{$^2$Department of Physics and Photon Science, GIST,
  Gwangju 61005, Korea}


\begin{abstract}
We present results of relativistic calculations of even order harmonic generation from
various atomic targets. The even order harmonics appear due to the relativistic non-dipole effects. 
We take these relativistic effects into account by using an approach based on the solution of the 
time-dependent Dirac equation. The spectra of the non-dipole even harmonics 
look qualitatively similar to the spectra of the dipole harmonics obeying the same classical
cutoff rule. The temporal dynamics of the formation of the non-dipole harmonics is, 
however, distinctly different from the process of
dipole harmonics formation. Even order harmonics emission is 
strongly suppressed at the beginning of the laser pulse, and the emission times of the non-dipole harmonics
are shifted  with respect to the bursts of the dipole emission. These features are
partly explained by a simple modification of the classical three-step model which takes
into account selection rules governing the emission of harmonic photons.

\end{abstract}

\maketitle


\section{Introduction}\label{seci}

One can expect relativistic effects to play important role in the dynamics 
of the processes of atomic or molecular interactions with strong laser pulses 
for laser intensities over $10^{18}$ W/cm$^{2}$ \cite{Reiss1998}, when, with increasing ponderomotive energy,
electron velocity can approach speed of light in the vacuum.
It has been realized since the pioneering paper by Reiss \cite{Reiss1989}, however, that
relativistic effects may reveal themselves even for moderately intense ($10^{13}$-$10^{14}$ W/cm$^{2}$) 
low frequency infrared (IR) laser fields. 
For instance, even for the IR laser fields of intensity of the order of $10^{13}$ W/cm$^2$,
the relativistic effects are visible in the photo-electron 
spectra \cite{ludw,recoil,ndi1,ndim,popov_rel,hatsa2,hatsa1} in the tunneling regime of ionization, 
characterized by the values  $\gamma \lesssim 1$, where 
$\gamma=\omega\sqrt{2|I_p}/E_0$ is the Keldysh parameter \cite{Keldysh64}, 
and  $\omega$, $E_0$ and $I_p$ are the field frequency, field 
strength and ionization potential
of the target system expressed in atomic units.
These relativistic non-dipole effects are due to the 
influence of the magnetic field component of the laser pulse
which induces a non-negligible momentum transfer to the photoelectrons 
\cite{Smeenk2011,Ludwig2014}. 
Alternatively, if we prefer the photon picture of light, one might say
that an IR photon carries small momentum, but a large number of
the photons participating in the process of the tunneling ionization \cite{kri}
deliver non-negligible momentum to the ionized electron \cite{chelnd1,chelnd2}.

The momentum delivered by the photons to the 
photo-electron was measured experimentally under 
the typical parameters of the tunneling ionization regime \cite{recoil0}. 
This momentum manifests itself, on average,  as a shift of the photo-electron momentum distributions 
(PMD) in the pulse propagation direction. More detailed picture, which emerges as a result of the 
complex interplay of the magnetic and Coulomb forces, includes 
the so-called direct  electrons which never recollide with the parent ion and are 
driven in the direction of the laser photon momentum, and the slow electrons which experience recollisions 
and may acquire 
momentum opposite to the photon momentum \cite{ndi1}.

Theoretical study of these effects clearly necessitates methods which 
go beyond the commonly used 
non-relativistic dipole approximation. A number of theoretical procedures allowing to 
consider the relativistic non-dipole effects have been described in the literature, including
the relativistic strong-field approximation \cite{Reiss2013,Yakaboylu2013,hatsa2,hatsa1},
time-dependent Schr\"odinger equation (TDSE) with non-dipole corrections 
\cite{chelnd1,chelnd2,chelnd3,ndim}, an approach based on the non-dipole strong-field-approximation Hamiltonian
\cite{madsennd}, and 
the time-dependent Dirac equation (TDDE) \cite{tdde1,lind1,relmine,spin_fl}.

The non-dipole effects manifest themselves as well in other processes occurring when atoms or molecules
interact with laser fields. The process which will interest us in the present work is the process of the
High Harmonic Generation (HHG). The non-dipole effects are known to produce several modifications
in the HHG spectra. It was found \cite{ndic} that the non-dipole interactions lead to 
decrease of harmonic intensity and 
shift of odd order harmonics in the spectra. A 
detailed investigation  of the effect of the pulse magnetic field on harmonic spectra 
was reported in 
\cite{ndic1,ndic2,ndic3}. It was found \cite{ndic1} that the non-dipole magnetic field effects result in 
the emission of photons polarized along the 
propagation direction which, for the laser pulse wavelength of $800$ nm and intensity 
of the order of $5\times 10^{15}$ W/cm$^2$, is 
several orders of magnitude weaker than the photon emission polarized parallel to the driving pulse 
polarization direction.  For stronger pulses with the intensities of the order of 
$10^{17}$ W/cm$^2$, the magnetic field effects start playing crucial role \cite{ndic3}.
Electron drift in the laser propagation direction due to the magnetic-field component of the laser pulse 
prevents recollisions, and hence, as one could expect on the basis of the picture provided by the 
celebrated three-step model of
HHG \cite{hhgd,Co94},  leads to the decrease of the harmonic emission. 

Perhaps one of the most striking manifestations of the non-dipole effects is
appearance of even order harmonics in the HHG spectra \cite{ndic6,ndic7,ndic5},
presenting an example of a relatively small perturbation producing not only relatively minor
quantitative modifications of the spectra, but introducing a qualitative change: harmonics with frequencies
forbidden in the dipole approximation. The appearance of the even order harmonics can be understood as a 
result of the break-up of the well-known symmetry which the electron trajectories responsible for the 
emission of the harmonic photons exhibit in the dipole approximation \cite{hhgd}. 
Magnetic field effects break this symmetry, 
and thus make possible generation of even order harmonics.
These harmonics were studied theoretically in \cite{ndic7}, 
using perturbative treatment of the non-dipole effects.

In the present paper we report a systematic theoretical study of the non-dipole effects, in particular 
generation of the even order harmonics, from various atomic targets. We use the TDDE as our main 
calculational tool, basing on the previously developed procedure for the numerical solution of the time-dependent 
Dirac equation \cite{relmine,spin_fl}. The approach based on the TDDE provides a 
complete non-perturbative description of the 
non-dipole, as well as other relativistic effects. 

Atomic units with $\hbar=1$, $e=1$, $m=1$, and $c \approx 137.036$  
(here $e$ and $m$ are charge and mass of the electron, 
$c$- speed of light) are used throughout the paper.

\section{Theory}\label{sect}

\subsection{Numerical solution to the time-dependent Dirac equation}

We solve the TDDE:

\begin{equation}
i\frac{\partial \Psi(r,t)}{\partial t}=\hat{H} \Psi(r,t)
\label{d}
\end{equation}

following the procedure we 
described in \cite{relmine,spin_fl}, which we briefly recapitulate below for the 
readers convenience.  In \Eref{d} 
$\Psi(r,t)$ is a four-component bispinor and the Hamiltonian operator has a form:

\begin{equation}
\hat H = \hat H_{\rm atom}+ \hat H_{\rm int}
\label{h}\ ,
\end{equation}

\noindent with:

\begin{equation}
\hat H_{\rm atom}=c{\bm \alpha}\cdot {\hat{\bm p}}+
c^2(\beta-I)+ I\ V(r)\ ,
\label{hatom}
\end{equation}

\noindent and

\begin{equation}
\hat H_{\rm int}=c{\bm \alpha}\cdot {\hat {\bm A}} \ ,
\label{hint}
\end{equation}

In \Eref{hatom}: 

$\displaystyle {\bm \alpha}=\left( \begin{array}{cc}
{\bm 0} & {\bm \sigma} \\
 {\bm \sigma} & {\bm 0}  \\
 \end{array} \right)$, $\displaystyle \beta=\left( \begin{array}{cc}
{\bm I} & {\bm 0} \\
{\bm 0} & -{\bm I}  \\
 \end{array} \right)$, $\displaystyle I=\left( \begin{array}{cc}
{\bm I} & {\bm 0} \\
{\bm 0} & {\bm I}  \\
 \end{array} \right)$, ${\bm \sigma}$ are Pauli matrices, 
${\bm 0}$ and ${\bm I}$ are $2\times 2$ null and identity matrices,  $V(r)$ is the atomic potential and $c=137.036$- the speed of light. We subtracted from the field-free atomic Hamiltonian \eref{hatom} 
the constant term
$Ic^2$ corresponding to the rest mass energy of the electron.

We use a laser pulse linearly polarized in $z-$ and propagating in $x-$ directions.
The vector potential of the pulse is defined in terms of the pulse electric field:

\begin{equation}
\bm{A}(x,t)= -\hat{\e_z}\int\limits_0^u E(\tau)\ d\tau  \ ,
\label{ef}
\ee

where $u=t-x/c$. At any given point in space the pulse has a finite duration $T_1$ so that 
$E(\tau)$ in \Eref{ef} is non-zero only for $0 < \tau < T_1$. As targets, we will consider below a model atom with a short range (SR) Yukawa-type
potential $V(r)=-1.903 e^{-r}/r$, hydrogen atom, and helium atom described by means of
an effective potential \cite{oep1}. The target atom is initially in
the ground $s-$ state $|\phi_0\rangle $ with the ionization potential (IP) of 
0.5 a.u. for the hydrogen and Yukawa atoms and IP of 0.902 a.u. for the He atom.

The solution to \Eref{d} is expanded as a series in the basis bispinors:

\begin{equation}
\Psi(\mathbf{r},t)=
\sum\limits_{j\atop l=j\pm 1/2} \sum\limits_{M=-j}^{j} 
\Psi_{jlM}({\bm r},t),
\label{basis}
\end{equation}

where:

\be
\Psi_{jlM}({\bm r},t)=
\left( \begin{array}{c}
g_{jlM}(r,t)\Omega_{jlM}(\n) \\
f_{jlM}(r,t)\Omega_{jl'M}(\n)  \\
 \end{array} \right),
\label{bb}
\ee

and the two-component spherical
spinors are defined as $\displaystyle \Omega_{jlM}(\n)=
\left( \begin{array}{c} 
C^{jM}_{l\ M-{1\over 2} {1\over 2}{1\over 2}}Y_{l,M-{1\over 2}}(\n) \\
C^{jM}_{l\ M+{1\over 2} {1\over 2}-{1\over 2}}Y_{l,M+{1\over 2}}(\n) 
\end{array} \right)$, (here $C^{jM}_{lm{1\over 2}\mu}$ are 
the Clebsch-Gordan coefficients, $Y_{lm}(\n)$- 
spherical harmonics, and $\n=\r/r$). Parameters $l$ and $l'$ in 
\Eref{basis} must satisfy the relation $l+l'=2j$.

To take into account the non-dipole effects due to the spatial dependence of the laser fields,
vector potential \eref{ef} is expanded in a series of spherical harmonics
at every time-step of the integration procedure. 
Substituting expansion \eref{basis} and expansion for the vector potential in the TDDE \eref{d}, and
using well-known properties of spherical spinors \cite{akhie,ll4}, one obtains 
a system of coupled differential equations for the radial functions  $g_{jlM}(r,t)$ and 
$f_{jlM}(r,t)$ in \Eref{bb}. This system has been solved 
using a relativistic generalization of the well-known matrix iteration method (MIM)
\cite{velocity1}, which we described in detail in \cite{relmine}. 

Appropriate choice of the propagation technique is essential, as the Dirac equation, as
it is well-known, possesses some properties which are absent in the case of the non-relativistic 
wave-equation. These properties are due to the presence of the continuum of the negative energy states
in the Dirac Hamiltonian which makes the Dirac Hamiltonian unbounded from below. One problem which 
this fact entails is the well-known problem of the collapse to the negative energies continuum 
\cite{varcol}, which  may manifest itself when basis set methods are
used to construct approximations to the bound states of
the Dirac Hamiltonian \cite{varcol}. We avoid this problem, since we do not rely on the basis
set methods. Initial state of the system is prepared in our calculation 
by solving numerically the
eigenvalue equation for the field-free Dirac Hamiltonian employing shooting method.
A related problem is the so-called Zitterbewegung problem \cite{drozh}.
Presence of a superposition of the states with positive and 
negative energies implies that a solution to the TDDE should exhibit very fast oscillations 
with characteristic frequencies of the order of $c^2$. Such oscillations are indeed present
and we can reproduce them in the framework of our numerical 
procedure by using sufficiently small integration time-step $\Delta $ \cite{relmine}.
We had to use the time-step $\Delta $ of the order of $10^{-6}$ a.u. in \cite{relmine} to reproduce these
oscillations. Use of such small values for $\Delta $, if it were imperative, would make any
practical calculations impossible, of course. Fortunately, one can bypass this problem 
by using an appropriate time-propagation 
technique. We discussed this issue in greater detail in 
\cite{tunrm,relmine}. For readers convenience we present a core of the argument below. 
From the purely numerical point of view, presence of the fast oscillating terms in 
a system of the ordinary differential equations (ODE)  gives us an example of a numerically stiff 
system of ODE, i.e. a system in which vastly different time-scales are present. To solve
such a system of ODE we must use a stable integration method \cite{stiff}, which ensures 
that while the numerical solution does not reproduce very fast oscillations, 
it describes accurately the overall behavior of the true solution. The integration procedure that we use 
provides such a stability. We can illustrate this point using a simple example
of a stiff system of two ODE:

\be
i \dot\y=\A\cdot\y,
\label{st1}
\ee

with Hermitian matrix $\A=diag(\lambda_1(t),\lambda_2(t))$. To mimic the problem at
hand let us assume that $\lambda_1$ is of order of $1$, while $\lambda_2$ has large negative value
on the interval of time that we consider. Short-time propagator in the MIM method is a 
unitary Crank-Nicholson (CN) propagator  \cite{crank}, which relates solution vectors
$\y_{n+1}=\y(t_{n+1})$ and $\y_n=\y(t_n)$ at times $t_n$ and $t_{n+1}=t_{n}+\Delta$ as follows:

\be
{\y}_{n+1}={1-{i\Delta\over 2}\A(t_{n+1/2})\over 1+{i\Delta\over 2}\A(t_{n+1/2})} {\y}_{n} \ ,
\label{st2}
\ee

where $t_{n+1/2}= t_n + \Delta/2$. 
One can see from \Eref{st2} that if at the $n-$th step of the propagation the second component 
of the vector $\y$ acquires a numerical error $\delta y^{(2)}_n$, the unitarity of the 
CN propagation matrix in \Eref{st2} makes this error remain bounded for 
$m>n$. 

Spatial variables in the coupled differential equations for the 
radial functions  $g_{jlM}(r,t)$ and 
$f_{jlM}(r,t)$  were discretized on a grid with the step size $\delta r=0.05$ a.u., the radial
variable was restricted to an interval $(0,R_{\rm max})$, with
$R_{\rm max}=400$ a.u., and angular momenta $j$ up to 70 were included in the expansion
\eref{basis} in the calculations below. The propagation time-step $\Delta$ was $0.05$ a.u.
Before proceeding to the description of  the results of this calculation, it
is instructive, however, to discuss an alternative treatment of the non-dipole
effects based on the leading order perturbation theory (LOPT) expansion,
as it provides a more transparent physical picture of the non-dipole effects 
than the complete Dirac equation.
LOPT calculation described below was also used as an accuracy  test for our solution to the 
TDDE.

\subsection{LOPT treatment of the non-dipole effects.}

We are interested in a LOPT solution to the TDDE considering the non-dipole
effects as relativistic corrections. 

The leading order relativistic  corrections describing the non-dipole effects in 
atom-field interaction can be obtained by expanding the minimal coupling 
atom-field interaction Hamiltonian \cite{lampe,Sobelman72,ndi1}
in the velocity gauge:

\be
\hat H^{\rm min}_{\rm int}(t)= \hat\p\cdot\A(\r,t)+ {\hat\A^2(\r,t)\over 2}
\label{min}
\ee

in powers of $c^{-1}$ \cite{ndim}:

\be
\hat H_{\rm min}(t) = \hat p_z A(t)+ {\hat v_z x E(t)\over c} 
+ {A^2(t)\over 2} + O(c^{-2})\  ,
\label{gauge}
\ee

where $\displaystyle E(t)=-{\partial A(t)\over \partial t}$ is the electric field of the pulse,
and the velocity operator $\displaystyle \hat \ve= \hat\p+ {\bm A}(t)$ has been introduced.
The last term on the r.h.s. of \Eref{gauge} is a function of time only and can be removed by a unitary 
transformation of the wave-function.  

Including  spin effects in the interaction Hamiltonian 
is not necessary, if we are interested in the effects of the leading order in powers of  
$c^{-1}$ \cite{ndi1,ndic}. The fact that the spin degrees of 
freedom can be neglected in the leading  
order of the $c^{-1}$ expansion, can be understood using the semi-classical picture of the 
spin effects, in which additional force due to the presence of the spin degrees of freedom, acting on the 
electron, is ${\bm F}=-\nabla U_m$, where  
$U_m= -{\bm \mu}\cdot{\bm H}$,  energy of the spin-magnetic field interaction. Here
${\bm H}$ is the magnetic field and ${\bm \mu}$ is electron's magnetic moment related
to the expectation value of electron's spin
${\bm \mu}=-2{\bm S}/c$. Spatial gradient of ${\bm H}$ introduces an additional
factor of $c^{-1}$, making contribution of the force $\bm F$  an effect of higher order in $c^{-1}$. 
As for the relativistic
corrections to the field-free atomic Hamiltonian, the so-called Breit-Pauli 
Hamiltonian \cite{Sobelman72}, it adds 
terms  of the order of $c^{-2}$  to the non-relativistic atomic Hamiltonian. 
We do not have, therefore, to include these corrections in the LOPT treatment.
To the leading order in powers of the $c^{-1}$-expansion, the dynamics of the system can thus be described by the 
time-dependent Schr\"odinger equation (TDSE):

\begin{equation}
i {\partial \Psi(\r,t) \over \partial t}= \left(\hat H_{\rm atom} +\hat H_{\rm d}(t)
+\hat H_{\rm nd}(t)\right)
\Psi(\r,t) \ ,
\label{tdse}
\end{equation}

where

\be 
\hat H_{\rm atom}=  {\hat\p^2\over2} +V(r) 
\label{hat}
\ee

is atomic field-free Hamiltonian,  

\be
\hat H_{\rm d}(t)= \hat p_z A(t) 
\label{hdip}
\ee

is the dipole part of the atom-field interaction and

\be
\hat H_{\rm nd}(t)  =  {\hat v_z x E(t)\over c} 
\label{hndip}
\ee

is the non-dipole part of the atom-field interaction containing the 
effects of the order of $c^{-1}$.

It is easy to check that the LOPT 
solution to the equation  \eref{tdse}, with the non-dipole term \eref{hndip} considered as a perturbation, can be written as:

\be
\Psi^{\rm LOPT}(\r,t)= \Psi_{\rm d}(\r,t) + \Psi^{(1)}_{\rm nd}(\r,t) \ ,
\label{pt}
\ee

where the LOPT non-dipole correction is given by the expression:

\be
\Psi^{(1)}_{\rm nd}(\r,t) = 
-i\int\limits_0^t \hat U_{\rm d}(t,\tau) 
\hat H_{\rm nd}(\tau) \Psi_{\rm d}(\r,\tau)\ d\tau  \ .
\label{ptn}
\ee

As can be seen from \Eref{hndip} for the operator $\hat H_{\rm nd}$
this correction is of the order of $c^{-1}$.
In \Eref{pt} and \Eref{ptn} $\Psi_{\rm d}(\r,t)$ is the zero-order solution to the 
non-relativistic TDSE taking into account only the dipole part of the 
atom-field interaction,
$\hat U_{\rm d}(t,\tau) $ is the evolution operator describing evolution of the system
driven by the non-relativistic dipole Hamiltonian. $\hat U_{\rm d}(t,\tau) $ 
satisfies the operator equation:

\be
i {\partial \hat U_{\rm d}(t,\tau)\over \partial t} = \left(\hat H_{\rm atom}+\hat H_{\rm d}(t)\right) 
\hat U_{\rm d}(t,\tau) \ ,
\label{evol}
\ee

and the initial condition $\hat U_{\rm d}(\tau,\tau)=\hat I$. 
In practice, we need not solve the 
operator equation \eref{evol}. All we have to do to compute the expression under the 
integral on the r.h.s of \Eref{pt} for given $\tau$ and $t$,
is to propagate first the initial state wave-function on the interval $(0,\tau)$ using the 
non-relativistic TDSE with the Hamiltonian \eref{hdip}, obtaining thus
a state vector $\Psi_{\rm d}(\tau)$. We act than on this vector with the operator 
$\hat H_{\rm nd}(\tau) $ and propagate it further in time till the moment $t$. 
The non-relativistic TDSE was solved using the well-tested 
numerical procedure described in \cite{cuspm}.

\subsection{Calculation of electron velocity and HHG spectra}

Once the solution to the TDDE \eref{d} is obtained,  expectations value of the electron 
velocity can be obtained as \cite{avet}:

\be
\ve(t)= c\langle \Psi(t)|{\bm \alpha}|\Psi(t) \rangle  \ .
\label{veld}
\ee

Harmonic spectra can then be calculated using the usual semi-classical approach, in which the 
spectral intensity
of the harmonic emission can be expressed in terms of the Fourier transform of electron's velocity:

\be
S_a(\Omega) \propto \left| \int\limits_0^{T_1}  v_a(t) W(t) 
e^{i\Omega t} \ dt \right|^2 \ .
\label{hhg}
\ee

where $v_a(t)$ is either $x-$ or $z-$ component of the electron velocity for the 
non-dipole and dipole harmonic intensities $S_x(\Omega)$ and $S_z(\Omega)$, 
respectively. In the velocity form for the harmonics intensity which we use here,
we do not need to introduce additional 
powers of harmonic frequency, which would be present had we used length or acceleration forms
\cite{madhhg}. The factor $W(t)$ in \Eref{hhg} is the window function \cite{agnhhg}, for which we 
employ the Hann form: $\displaystyle W(t)= \sin^2{\left(\pi t\over T_1\right)} $.

The most noticeable effects which the relativistic non-dipole corrections produce are 
appearance of harmonic photons polarized in the laser propagation direction 
\cite{ndic1,ndic2,ndic3} and appearance of even order harmonics
in the HHG spectra \cite{ndic4,ndic5}. The LOPT picture allows to explain these features transparently.
Substituting the expression \Eref{pt} for the LOPT wave-function into the matrix element:  

\be 
\langle\Psi^{\rm LOPT}(t)|\hat\ve|\Psi^{\rm LOPT}(t)\rangle \approx
\hat{\bm x}v_{x}(t) + \hat{\bm y}v_{y}(t) + \hat{\bm z}v_{z}(t) \ ,
\label{dd}
\ee

defining the leading order contributions to the 
expectation value of electron velocity, one obtains:

\be
v_{z}(t) = \langle\Psi_{\rm d}(t)|\hat v_z|\Psi_{\rm d}(t)\rangle \ .
\label{velz}
\ee

For the geometry we use, the evolution operator $\hat U_{\rm d}(t,\tau)$ commutes with 
$\hat l_z$- the $z-$ component of the angular momentum, i.e., it is a conserved quantity for the 
quantum evolution driven by the dipole Hamiltonian \eref{hat} and \eref{hdip}.
${\hat l}_z$, therefore, 
has a definite value $l_z=0$ in the state described by the wave-function $\Psi_{\rm d}(t)$,
and the matrix element $\displaystyle \langle\Psi_{\rm d}(t)|\hat v_x|\Psi_{\rm d}(t)\rangle $
vanishes because of the well-known dipole selection rules \cite{Sobelman72}. Leading order
contribution to $v_x(t)$, is, therefore, of the order of $c^{-1}$, and is given by the expression:

\ba
v_{x}(t) &=& \langle\Psi_{\rm d}(t)|\hat v_x|\Psi^{(1)}_{\rm nd}\rangle + 
\langle\Psi^{(1)}_{\rm nd}(t)|\hat v_x|\Psi_{\rm d}\rangle \nonumber \\
&=& 2{\rm Re} \langle\Psi_{\rm d}(t)|\hat v_x|\Psi^{(1)}_{\rm nd}\rangle \nonumber \\
&=& 2{\rm Im}\left( 
\int\limits_0^t \langle \Psi_{\rm d}(t)|\hat p_x \hat U_{\rm d}(t,\tau) 
\hat H_{\rm nd}(\tau) |\Psi_{\rm d}(\tau)\rangle \ d\tau \right) \nonumber \ . \\
\label{velx}
\ea

In the last line of \Eref{velx} we used expression \eref{ptn} for $\Psi^{(1)}_{\rm nd}$.
The same dipole selection rules \cite{Sobelman72} and the structure of \Eref{ptn} ensure that the contribution  
of the order of $c^{-1}$ to $v_y(t)$ is zero. 
The leading contribution of the non-dipole effects is, therefore, 
non-zero only for the $x$-component of the electron velocity.
Orientation of the dipole velocity due to this relativistic contribution results, thus,
in the appearance of the harmonic photons polarized in the propagation direction in accordance with
the observations made in \cite{ndic1,ndic2,ndic3}.

As we mentioned above, the appearance of the even order harmonics can be understood as a
result of violation of the symmetry of the electron trajectories responsible for the 
emission of harmonic photons in the dipole approximation \cite{hhgd}. From the LOPT 
perspective this effect can be explained as follows. 
As one can see from \Eref{hdip} and \Eref{hndip}, the dipole interaction 
operator \eref{hdip} has odd parity, i.e. it couples 
states of different parities, while the non-dipole operator \eref{hndip} has even parity. 
Employing a somewhat
lousy language, we might say that the presence of these two atom-field interaction Hamiltonians
can be described as the presence of two kinds of photons: 
the "dipole" photons and the "non-dipole" 
photons, whose emission and absorption are governed by the operators \eref{hdip} and 
\eref{hndip}, respectively.  Using these notions and the  LOPT expression for
$v_{x}(t)$ in \Eref{velx}, contribution of the 
non-dipole interaction to the formation of the $N-$th harmonic can be described as absorption of 
$N-1$ "dipole" photons and one "non-dipole" photon, with subsequent recombination to atomic ground state accompanied by emission of a harmonic photon with frequency $N\omega$. Using the informal terminology 
which we adopted, one might say that the emitted harmonic photon is of the 
"dipole" nature since spontaneous emission 
satisfies the dipole selection rules. Conservation of the total parity for the combined 
system of atom and the "dipole" and the "non-dipole" photons implies then 
that $N$ must necessarily be even.

Besides  providing a simple physical picture of the appearance of even harmonics, the LOPT
approach which we described above, can be used as a test of the accuracy of our solution to the 
TDDE. To do such a test we performed calculations of the expectation values of electron velocity using 
TDDE and LOPT approaches for the cosine-pulse form shown in \Fref{f1}, with the 
vector potential in \Eref{ef} given by the equation: 
$\displaystyle \A(x,t)= -\e_z{E_0\over \omega} \sin^2{\left(\pi u\over T_1\right)} \sin{\omega u}$ 
where $\omega=0.057$ a.u., $E_0=0.0534$ a.u., $u=t-x/c$.  A comparison of the TDDE results obtained using 
\Eref{veld} and the LOPT results obtained using \Eref{velx} for 
the $x-$ component of electron velocity is shown in \Fref{f2}. 
The results of the LOPT
treatment prove to be virtually identical to the results of the TDDE calculation which 
is not surprising given that the relativistic corrections could be expected to 
be small for the field parameters we consider.

\section{Results}\label{sec2}

We report below results which we obtained from our TDDE calculations for
dipole $S_z(\Omega)$ and non-dipole $S_x(\Omega)$ harmonic intensities 
for different targets. HHG spectra were obtained by computing electron velocity 
as prescribed by \Eref{veld} and using \Eref{hhg} to compute harmonic intensities.
Calculations were performed using the sine waveform shown in \Fref{f1} with 
the electric field given by the equation: 
$\displaystyle E(u)= E_0 \sin^2{\left(\pi u\over T_1\right)} \sin{\omega u} $.
We report below results for the base frequencies $\omega=0.114$ a.u. (wavelength of $400$ nm)
and $\omega=0.057$ a.u. (wavelength of $800$ nm).

In \Fref{f3} we show HHG spectra that we obtained for the driving pulse wavelength 
$\lambda= 400$ nm and different field strengths for various targets. \Fref{f3} shows 
both dipole $S_z(\Omega)$ and non-dipole $S_x(\Omega)$ harmonic intensities. The vertical lines in
the Figures show positions of the classical cutoffs given  by the well-known
$3.17U_p+ I_p$ (here $U_p=E_0^2/4\omega^2$ and $I_p$ are ponderomotive and ionization 
energies respectively) rule of the three-step model \cite{hhgd,Co94}. 
In \Fref{f4} we zoom on the parts of the harmonic spectra more closely to demonstrate the presence of  
odd and even harmonics in the dipole and non-dipole spectra respectively. 

Quite expectedly, behavior of the dipole intensity $S_z(\Omega)$ shown in \Fref{f3} 
agrees very well
with the three-step model predictions, exhibiting a sharp drop in magnitude after 
reaching the classical cutoff. The non-dipole $S_x(\Omega)$ spectra mimic
this behavior very closely. This may be not surprising if we make use again of 
the LOPT picture of 
formation of the non-dipole harmonics we presented above, which relied on the notions 
of 'dipole' and 'non-dipole' photons with operators describing their interactions with 
an atom given by \Eref{hdip} and \Eref{hndip}, respectively. We remind, that 
in the framework of this picture the $N-$th non-dipole harmonic is produced 
as a result of the absorption of $N-1$ "dipole" photons and one "non-dipole" photon.
As far as the harmonic spectra are concerned, 
the mechanism responsible for the formation  
of the non-dipole harmonic emission differs thus from the 
mechanism of the emission of the dipole harmonics only in the replacement of one 'dipole' photon
with a 'non-dipole' one. This replacement leads to the replacement of the 
odd order harmonics in the spectra by the even order ones and results in
an overall drop in magnitude
in the harmonic spectra due to the presence of the additional factor of $c^{-1}$ in the non-dipole interaction 
operator \eref{hndip}. 

The energy and parity conservation considerations which lead us to the general conclusions about the 
character of the 
non-dipole spectra do not tell us anything about temporal dynamics of the formation of the 
non-dipole harmonics. We can have a glimpse of this temporal dynamics by
analyzing Gabor transforms \cite{gabt} of dipole and non-dipole velocities:

\be
T_a(\Omega,t)= \int\limits_0^{T_1} v_a(\tau) \Phi^*(t,\tau,\Omega) d\tau \ ,
\label{wav}
\ee

where $\displaystyle \Phi(t,\tau,\Omega)= \exp{\left\{i\Omega\tau - (t-\tau)^2/2(x_0T)^2\right\}} $,
parameter $x_0$ determines resolution in the temporal domain, and $T$ is an optical cycle of the 
laser field. Gabor transform, as well as closely related wavelet transform, 
allows us to take a look simultaneously at both time
and frequency domains, and allows to determine, in particular,  when different 
harmonics are emitted \cite{wavelet4,wavelet,waveletc}. We used $x_0=0.1$ in the calculations 
below. This value of $x_0$ gives us rather poor resolution in the frequency domain, but
high resolution in the time domain, which is of interest to us presently.

The absolute values $|T_a(\Omega,t)|$ for both dipole and non-dipole velocities are shown in 
\Fref{f5} and \Fref{f6} for the SR Yukawa and hydrogen atoms. One can see that, dynamically, 
formation of dipole and non-dipole harmonics proceeds quite differently. 
For both Yukawa and hydrogen atoms systems emission of the non-dipole harmonics is 
strongly suppressed at the early stages of 
pulse development, and emission times for the non-dipole harmonics are shifted with respect to 
the dipole radiation bursts. Such behavior could be anticipated by looking at  \Fref{f2} 
which shows that $x-$ component of the velocity starts actually respond to the field 
only for times approaching the midpoint of the pulse. The reason for this could be
traced back to the character of the fully quantum expression for the velocity component $v_x$ in the 
second LOPT equation \eref{velx}, with time integration on the right-hand side of this equation 
smoothing out high frequency oscillations. 
To elucidate this issue further we performed  a simple classical calculation of the 
emitted photon energy as a function of the recombination time using the physical picture provided by the 
three-step model.  We assume that electron is ionized at the moment of time $t_{ion}$
and returns to the parent ion at the moment of time $t_{ret}$,
emitting a harmonic photon with energy $E_{ret}+ I_p$. As is usually assumed in the 
three-step model calculations, we consider only the effect of the external field \eref{ef} on the 
electron motion, neglecting completely ionic potential.
The only difference of our calculation and the traditional three-step 
model analysis of the harmonic emission, is that we take into account effect of the 
Lorentz force due to the magnetic field of the 
pulse. We simulate electron motion in a plane (which is the $(x,z)$- plane for the geometry 
we employ), solving the set of the classical Newton equations, which for the fields configuration,
geometry and atomic units system we employ, can be written as:

\ba
\ddot{x}&=& -{{v}_z\over c} E(t) \nonumber \\
\ddot{z}&=& - E(t) + {{v}_x\over c} E(t) \nonumber \\ \ .
\label{new}
\ea

Following the prescription of the traditional three-step 
model we solve equations \eref{new} with zero initial conditions imposed at the 
ionization time: $v_x(t_{ion})=v_z(t_{ion})=0$ and $x(t_{ion})=z(t_{ion})=0$.
We assume that the electron trajectory returns to the origin, if at the moment of time
$t_{ret}$,  $z-$coordinate of the electron trajectory changes sign.

\Fref{f7}(a) shows results of such a simulation, which qualitatively agree with the dynamics 
of the dipole harmonics emission shown in \Fref{f5} and \Fref{f6}, with bursts of harmonics emission occurring
every half cycle of the laser pulse. To be able to apply this classical analysis
to the emission of the non-dipole harmonics we must, however, introduce one essentially 
quantum ingredient in the model described by the classical equations\eref{new}. 
Emission of the non-dipole radiation differs from the 
emission of the dipole harmonics in one
important aspect. For the geometry we employ, the dipole harmonics photon emission process 
satisfies selection rule $\Delta M=0$,
where $M$ is the $z-$ projection of the electron angular momentum. On the other hand, emission
of the non-dipole harmonic photon, as can be seen from the LOPT analysis we presented above, 
must satisfy selection rule $\Delta M=\pm 1$. This means that for the ground $s-$ state that
we consider, non-dipole radiation can be emitted only by electrons with non-zero angular momentum.
We can incorporate this fact in our classical model by introducing a filter parameter 
$f$ in the simulations, and considering only those returning trajectories  
for which at the moment of time $t_{ret}$ squared classical angular momentum value exceeds the
threshold value set by the filter parameter $f$.
Results of such calculations are shown in \Fref{f7}(b-d) for different values of the filter parameter 
$f$. One can see that by increasing the value of the filter parameter, we make the classical picture
in \Fref{f7} look more like the Gabor transform results shown in \Fref{f5} and \Fref{f6}.
In particular, \Fref{f7}(c-d) show the absence of the non-dipole harmonics emission during the 
first two cycles of the laser pulse, the feature which is also demonstrated by the quantum 
analysis based on the Gabor transform in \Fref{f5} and \Fref{f6}. Applying non-zero
filter parameter does not change, however, the maximum energy $E_{ret}$ of the returning electron, 
which explains why non-dipole harmonic emission spectra exhibit essentially the same cutoffs as 
the dipole harmonic emission spectra.
This simple classical picture of
the formation of the non-dipole harmonics, which takes as quantum ingredient only the requirement
that the electron angular momentum on the returning trajectories should exceed certain threshold value, 
agrees, thus, qualitatively with the fully quantum picture.

We also performed TDDE calculations for the pulse base frequency 
$\omega=0.057$ a.u. (corresponding to the wavelength of $800$ nm).
In \Fref{f8} and \Fref{f9} we show harmonic spectra
we obtain from TDDE for the SR Yukawa and hydrogen atoms. \Fref{f10} shows
results of the analysis of the temporal dynamics of the harmonic formation 
based on the Gabor transform \eref{wav}. These Figures show essentially the same picture
as the results we presented above for the driving pulse wavelength of $400$ nm. The spectra 
of the non-dipole harmonics follow closely the classical dipole cutoff rule, and differ in 
this respect from the dipole emission spectra only in their intensity. Temporal 
pictures of the harmonics formation in the dipole and the non-dipole cases are, however, 
totally different. The main difference is, just as in the case of the driving pulse 
wavelength of $400$ nm, the absence of the harmonic emission at the early stages of the pulse development, the 
feature which we explained above using the results of the classical calculations shown 
in \Fref{f7}(c,d).

The factor which is responsible for the difference in intensity between the dipole and non-dipole 
harmonics is the additional factor of $c^{-1}$ which, as one can see from 
\Eref{velx} and \Eref{hndip}, is present in the LOPT formula for the $x-$component of the 
velocity. The presence of this factor in $v_x$ leads to a dampening factor of 
$c^{-2}$ in the expression for the non-dipole harmonics intensity. 
It is rather difficult to obtain a more detailed insight about relative magnitude of the 
dipole and non-dipole harmonic intensities form the cumbersome LOPT expressions
\Eref{velz} and \Eref{velx}. One can, however,  obtain a simple estimate 
using the reasoning based not on the Schr\"odinger picture that we have used so far, but
on the equivalent Heisenberg picture of the quantum mechanics (QM). In the latter, we remind, the 
operators evolve in time, while the state vectors do not. We obtain, of course, the 
same expectation values for all physical observables in both pictures. 

In the 
Heisenberg picture time-evolution of the operators ${\hat \r}(t)$ and ${\hat \p}(t)$ is described by the equations
\cite{LL3}:
$\displaystyle i{\dot{\hat\r}}= [\hat{\r},\hat H]$, $ \displaystyle i{\dot{\hat \p}}= [{\hat \p},\hat H]$,
where the Hamiltonian operator in our problem 
is $\hat H= \hat H_{\rm atom}+ \hat H_{\rm d}(t)+\hat H_{\rm nd}(t)$,
with $\hat H_{\rm atom}$,  $\hat H_{\rm d}(t)$ and $\hat H_{\rm nd}(t)$ given by \Eref{hat},
\Eref{hdip}, and \Eref{hndip}, respectively. Calculating the commutators, one obtains 
the following equations of motion:

\ba
\dot{\hat x}&=& \hat{p}_x \nonumber \\
\dot{\hat p}_x&=& -i[{\hat p}_x, \hat V] - {\hat{v}_z\over c}E(t) \nonumber \\
\dot{\hat z}&=& \hat{p}_z +A(t)+ {{\hat x}\over c}E(t) \nonumber \\
\dot{\hat p}_z&=& -i[{\hat p}_z, \hat V] \nonumber \ ,\\
\label{he}
\ea

where $\hat V$ is atomic potential operator, 
$\hat{v}_z= \hat{p}_z + A(t)$, $A(t)$ and $E(t)$ are the vector potential and the electric field of
the pulse. \Eref{he} is the quantum-mechanical analogue of the classical equations describing 
electron motion in the potential $V$ in presence of the Lorenz force. 
It contains the same physical 
information and is, therefore, equivalent to the 
LOPT equations \Eref{velz} and \Eref{velx}, but it provides
a more clear physical picture and can be used as a 
starting point for making simplifying assumptions. 

From the first two equations \eref{he} one obtains:

\be
\ddot{\hat x}= -i[{\hat p}_x, \hat V]- {\hat{v}_z\over c}E(t) \ ,
\label{lor}
\ee

We will make an assumption that one can omit the commutator
$[{\hat p}_x, \hat V]$ in \Eref{lor}. Some justification for this operation 
can be provided in the case of the SR Yukawa atom, when potential function $V(\r)$ is effectively zero 
everywhere excepting a small neighborhood of the atom. We obtain then
from \Eref{lor} a relation for the expectation values of the electron acceleration 
$\displaystyle a_x=\langle\phi_0|\ddot{\hat x}|\phi_0\rangle$ and velocity
$\displaystyle v_z=\langle\phi_0|{\hat v}_z|\phi_0\rangle$:

\be
a_x= -{v_z\over c}E(t) \ ,
\label{ac}
\ee

where $|\phi_0\rangle$ is the initial atomic state, which does not evolve in time in the Heisenberg
picture. Assuming further that $E(t)$ is a monochromatic wave:
$E(t)=E_0\cos{\omega t}$ and calculating Fourier transforms of both sides of
\Eref{ac}, 
we obtain a relation between the Fourier transforms 
$\displaystyle \tilde v_x(\Omega)=\int v_x(t)e^{i\Omega t}\ dt$ and 
$\displaystyle \tilde v_z(\Omega)=\int v_z(t)e^{i\Omega t}\ dt$:

\be 
-i\Omega\tilde v_x(\Omega)= 
{E_0\over 2c}\left(\tilde v_z(\Omega+\omega)+ \tilde v_z(\Omega-\omega)\right) \ ,
\label{til}
\ee

from which, using the fact that for any complex numbers $z_1$, $z_2$: 
$\displaystyle |z_1+z_2|^2 \le (|z_1|+|z_2|)^2$, we obtain 
an inequality:

\be 
\Omega^2 S_x(\Omega) \le  {E_0^2\over 4c^2}
\large( \sqrt{S_z(\Omega+\omega)}+ \sqrt{S_z(\Omega-\omega)}\large)^2
\label{til1}
\ee

We see from \Eref{til1} that for $\Omega > \omega$ we have:

\be 
R(\Omega)= {4c^2 \omega^2\over E_0^2} 
{S_x(\Omega)\over \large( \sqrt{S_z(\Omega+\omega)}+ \sqrt{S_z(\Omega-\omega)}\large)^2}  \le 1 \ .
\label{til2}
\ee

Introducing the magnitude $A_0=E_0/\omega$ of the pulse vector potential, we can rewrite 
inequality \eref{til2} as:

\be 
{S_x(\Omega)\over \large( \sqrt{S_z(\Omega+\omega)}+ \sqrt{S_z(\Omega-\omega)}\large)^2} \le
{A_0^2\over 4c^2}
\label{til3}
\ee

The ratio $R(\Omega)$ defined in \Eref{til2}
is shown in \Fref{f31}(a) for the SR Yukawa potential and various pulse parameters. 
Of course, we cannot expect \Eref{til2} to provide a rigorous upper bound since 
deriving it we neglected atomic potential in \Eref{lor}, which constitutes 
a rather drastic approximation. As one can see from \Fref{f31}(a)
inequality \eref{til2} can indeed be violated.
One can see, nevertheless, that \Eref{til2}, 
and consequently \Eref{til3} provide reasonably accurate estimates of the relative magnitude of the 
intensities of the dipole and non-dipole harmonics.

While the non-zero expectation value $v_x$ and appearance of the non-dipole harmonics
is an entirely relativistic phenomena, the non-dipole effects also modify slightly 
the velocity component $v_z$. The magnitude of this effect 
is of the order of $c^{-2}$. This can be most easily seen from the 
Heisenberg equations of motion \eref{he}. The equation for 
$v_z(t)$ (the third of the equations \eref{he}) contains the term ${\hat x} E(t)/c$ on the 
right-hand side. Since the expectation value of $x$ is itself of the order of $c^{-1}$, the resulting
effect on $v_z(t)$ is of the order of $c^{-2}$, which  will produce a relativistic correction of 
the order of $c^{-2}$ for the dipole harmonic intensity. We may expect, therefore, that the normalized difference:

\be
{\Delta S_z(\Omega)\over S_z(\Omega)}=
{ S_z(\Omega)-S^{\rm nr}_z(\Omega)\over S^{\rm nr}_z(\Omega)} \ ,
\label{r1}
\ee

where $S_z(\Omega)$ is the dipole harmonics intensity obtained in the present TDDE calculation and 
$S^{\rm nr}_z(\Omega)$ is the result of the non-relativistic TDSE calculation,
should be of the order of $c^{-2}$. i.e., we may expect 
$\Delta S_z(\Omega)/S_z(\Omega) \sim 10^{-4}$. That this is indeed the case can
be seen from \Fref{f31}(b), where we show results of the TDDE and TDSE calculations
performed for the same pulse parameters for the Yukawa atom.

The analysis based on the Heisenberg equations of motion \eref{he} also allows to give a simple explanation 
for the behavior of $v_x(t)$ shown in \Fref{f2}, where the $x-$ component of the electron velocity starts 
responding to the field only for the times approaching the midpoint of the pulse. 
Integrating \Eref{ac} we obtain for the expectation value 
$\displaystyle v_x=\langle\phi_0|{\hat v}_x|\phi_0\rangle$ (assuming that it has zero value at $t=0$):

\be
v_x(t)= -{1\over c} \int\limits_0^t v_z(\tau)E(\tau)\ d\tau \ .
\label{vc}
\ee

We could have obtained the same equation by integrating the first of the set of the 
classical equations \eref{new}, which is not surprising given the great formal similarity between
the classical mechanics and the QM in the Heisenberg picture. 
We show in \Fref{f12}(a) the expectation value $v_z(t)$ obtained in the LOPT calculation
for the cosine pulse with $E_0=0.0534$ a.u. and  $\omega=0.057$ a.u.
We show only the LOPT result. Just as in the case of $v_x(t)$, shown in \Fref{f2}, the 
TDDE and LOPT results for $v_z(t)$ differ very slightly. In \Fref{f12}(b) we 
show the LOPT expectation value  $v_x(t)$, as well as the estimate for $v_x(t)$ that
we obtain if we substitute the LOPT value for $v_z(\tau)$ under the integral sign in 
\Eref{vc}. One can see that the estimate thus obtained reproduces fairly well the general behavior 
of  $v_x(t)$. In particular, it reproduces the feature that 
we mentioned above: the $x-$component of the velocity begins deviating from zero appreciably only for the
times approaching the midpoint of the pulse. 
We remind that effect of the atomic potential on the motion in the $x-$direction was
neglected in the Heisenberg equation of motion \eref{ac} which we used to obtain
the estimate \eref{vc}. The fact that the estimate \eref{vc} reproduces qualitative behavior
of the $x-$component of electron velocity shown in \Fref{f2} tells us, therefore, that this behavior 
is a result of the interplay of the motion in $x-$ and $z-$ directions which are mutually interconnected due to  
presence of the Lorentz force.

\section{Conclusion}

We have presented results of the relativistic calculations of even harmonic generation from
various atomic targets. Our approach was based on the numerical solution of the TDDE.
The HHG spectra of the non-dipole even order harmonics were found 
to look qualitatively similar to the spectra of the dipole harmonics, obeying the same classical
cutoff rules. The temporal formation of the non-dipole harmonics, however, was found to be 
quite different.
The results of the Gabor transform analysis show that formation of the non-dipole harmonics is
strongly suppressed at the beginning of the laser pulse, and bursts of the non-dipole radiation
are shifted in time with respect to the bursts of the dipole emission. These features are
partly explained by a simple generalization of the classical three-step model,
which takes into account the selection rules governing emission of harmonic photons. We modeled the 
effect of these
selection rules by using a filter parameter, which selects the trajectories with angular momentum 
exceeding a certain threshold value at the recollision time.

For the field parameters we considered the relativistic effects are still relatively weak and 
could be described perturbatively. LOPT provides, as we have seen, an 
adequate description of the non-dipole effects responsible for the even order harmonics emission.
Use of the TDDE, however, is technically simpler than the calculations based on the LOPT, and 
opens the perspective of making an excursion into 
the truly relativistic domain in the future. We relied, therefore, on the TDDE-based approach 
in the present work. The present approach can also be generalized relatively 
easily to include some quantum electrodynamical (QED) effects, such as 
the vacuum polarization effects, or the QED strong
Coulomb field radiative corrections, which can be taken into account by using effective
potentials such as the Uehling potential \cite{eff_qed1,eff_qed3} or the radiative potential proposed in 
\cite{eff_qed2}. The procedure we apply to solve the Dirac equation can also be used 
to study the process of electron-positron pair production (PP) in strong electromagnetic fields, which occurs when 
field strength reaches the characteristic Schwinger field strength of $1.3\times 10^{16}$ V/cm. 
The process of  PP in both homogeneous and inhomogeneous electric fields has received
considerable interest in the literature \cite{pair_prod_inh}. Theoretical treatment of  PP in the
semiclassical approximation relies on a solution of the TDDE for a given
field configuration \cite{pair_prod}. Our procedure 
might prove useful for this purpose, especially in the
case of the spatially inhomogeneous field, which has been found to play an important role in the PP
\cite{pair_prod_inh,pair_prod_inh1}.

The numerical procedure we employ 
can be relatively easily generalized for the case of the many-electron 
relativistic Hamiltonians used in the quantum chemistry calculations \cite{muld1,muld2}. 
Use of the representation of the wave-function analogous to the expansion 
\eref{basis} would be, of course, impractical for systems with more than one electron 
if we want to use such expansions to represent the wave-function in the whole 
space. One may use, however, the idea of the $R-$matrix approach, which 
separates the coordinate space in the inner region, where
a suitable basis set representation can be used to represent  many-electron wave-functions and 
the outer region, where one has to concentrate on the description of a single electron motion,
for which the finite difference method might be better suited. Such a strategy has been implemented with
success in the framework of the so-called R-Matrix incorporating Time method (RMT) \cite{Rmat}
which allows to solve the non-relativistic TDSE for many-electron systems. 
One can use a similar approach in the relativistic case, relying on the results of the 
stationary quantum chemistry calculations \cite{muld1,muld2} for the description of the 
inner region, where many-electron effects are important, and using the present procedure 
to solve the TDDE describing electron
propagation in the outer region.

\section*{Acknowledgments}

This work was supported by the Institute for Basic Science grant (IBS-R012-D1) and 
the National Research Foundation of Korea (NRF), grant funded by the Korea government (MIST) (No. 2022R1A2C3006025). 
Computational works for this research were 
performed on the IBS Supercomputer Aleph in the IBS Research Solution Center.

\begin{figure*}[h]%
\centering
\includegraphics[width=0.9\textwidth]{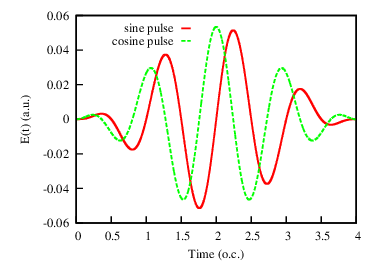}
\caption{(Color online) Pulse shapes $E(t)$ employed in the calculations.}
\label{f1}
\end{figure*}

\begin{figure*}[h]%
\centering
\includegraphics[width=0.9\textwidth]{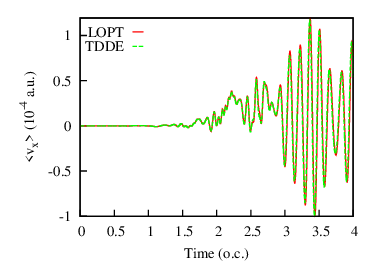}
\caption{(Color online) Expectation value of the $x-$ component of the electron velocity as a function 
of time obtained in TDDE and LOPT calculations 
Cosine pulse with $E_0=0.0534$ a.u., $\omega=0.057$ a.u.
has been used in the calculation.}
\label{f2}
\end{figure*}

\begin{figure*}[h]%
\centering
\includegraphics[width=0.9\textwidth]{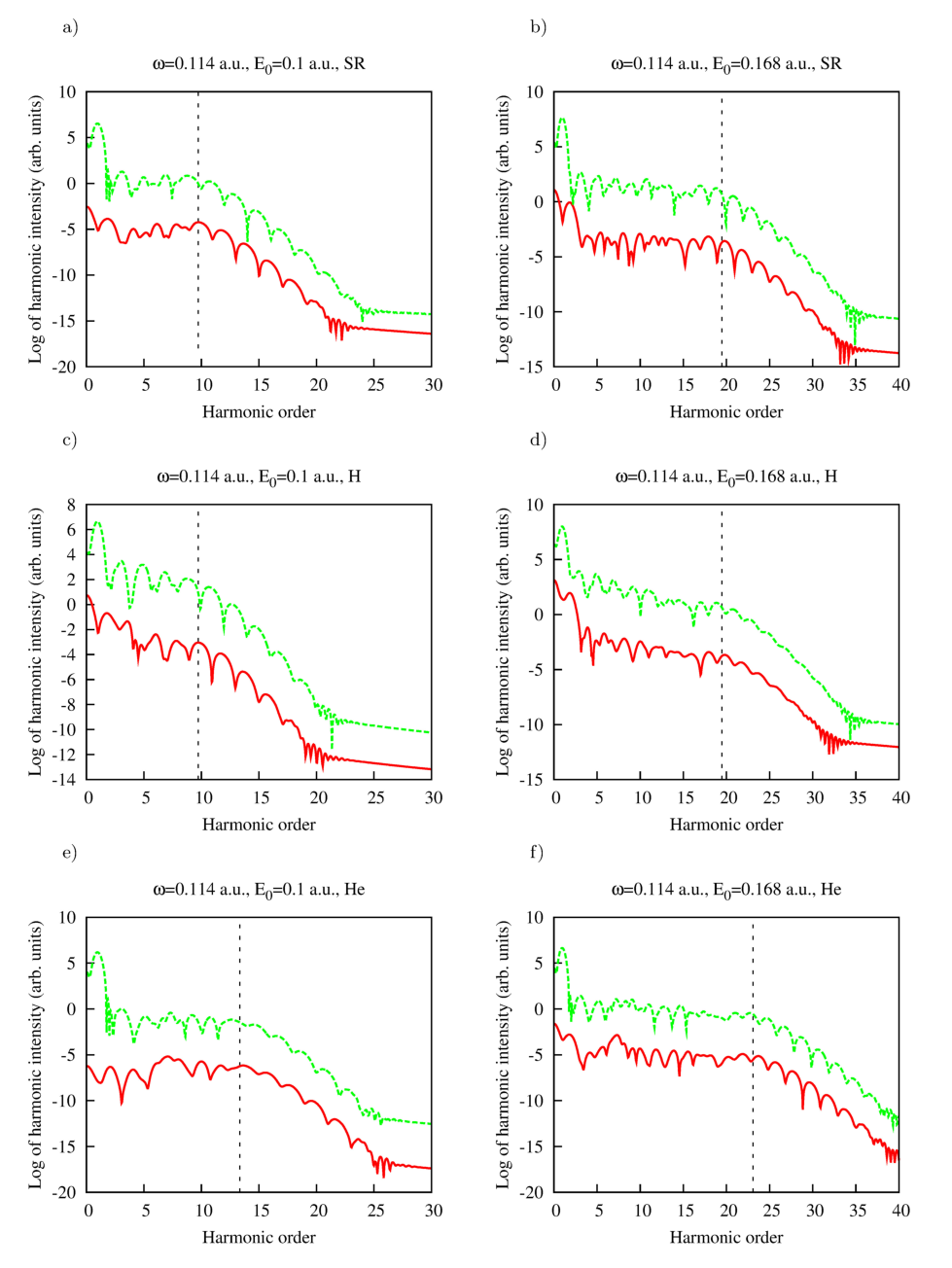}
\caption{(Color online) Dipole (dash green) and non-dipole (red solid) harmonic 
intensities for the pulse wavelength $\lambda=400$ nm 
for the SR Yukawa, hydrogen and He atoms.
Vertical dash lines show cutoff positions.
}
\label{f3}
\end{figure*}

\begin{figure*}[h]%
\centering
\includegraphics[width=0.9\textwidth]{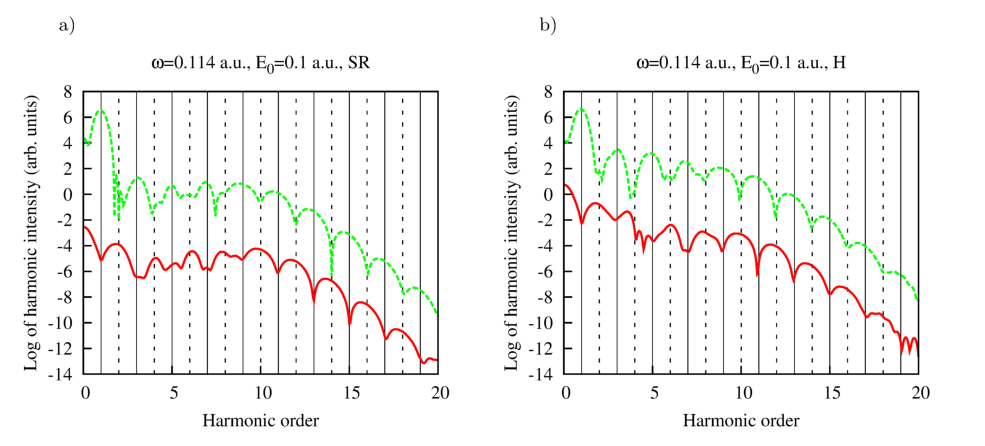}
\caption{(Color online) Dipole (dash green) and non-dipole (red solid) harmonic 
intensities for the pulse wavelength $\lambda=400$ nm for harmonics with orders 
$n \le 20$ for the SR Yukawa and hydrogen atoms.
Vertical solid and dash lines show positions of odd and even harmonics, respectively.
}
\label{f4}
\end{figure*}

\begin{figure*}[h]%
\centering
\includegraphics[width=0.9\textwidth]{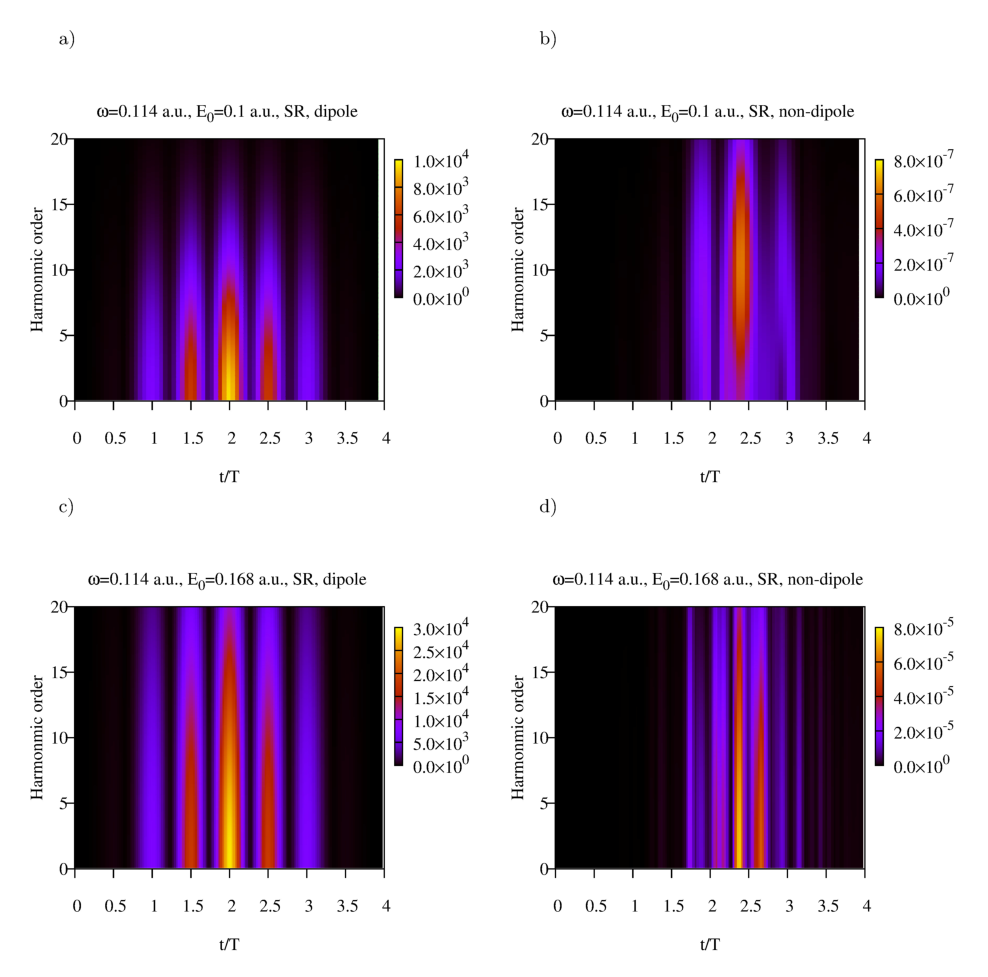}
\caption{(Color online) Gabor transform $|T(\Omega,t)|$ for 
the pulse wavelength $\lambda=400$ nm and different field strengths for 
the SR Yukawa atom.
}
\label{f5}
\end{figure*}

\begin{figure*}[h]%
\centering
\includegraphics[width=0.9\textwidth]{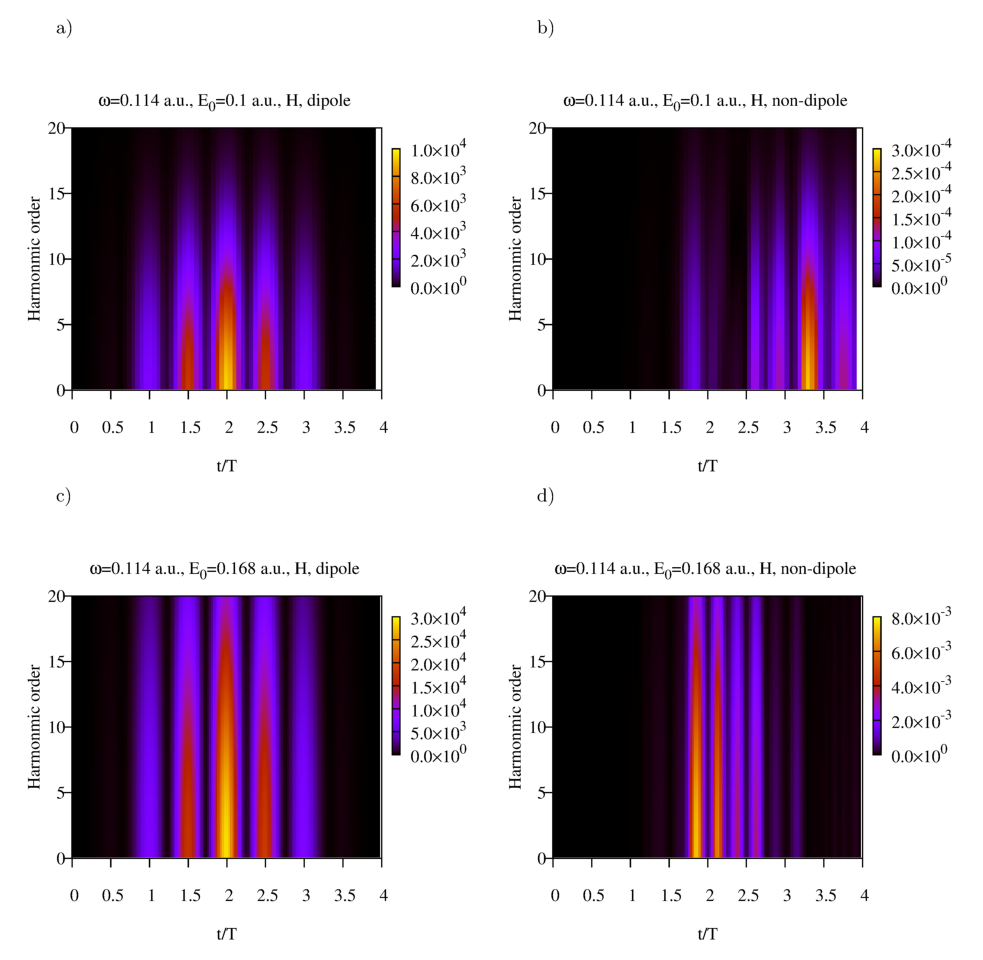}
\caption{(Color online) Gabor transform $|T(\Omega,t)|$ for 
the pulse wavelength $\lambda=400$ nm and different field strengths for the 
hydrogen atom.
}
\label{f6}
\end{figure*}

\begin{figure*}[h]%
\centering
\includegraphics[width=0.9\textwidth]{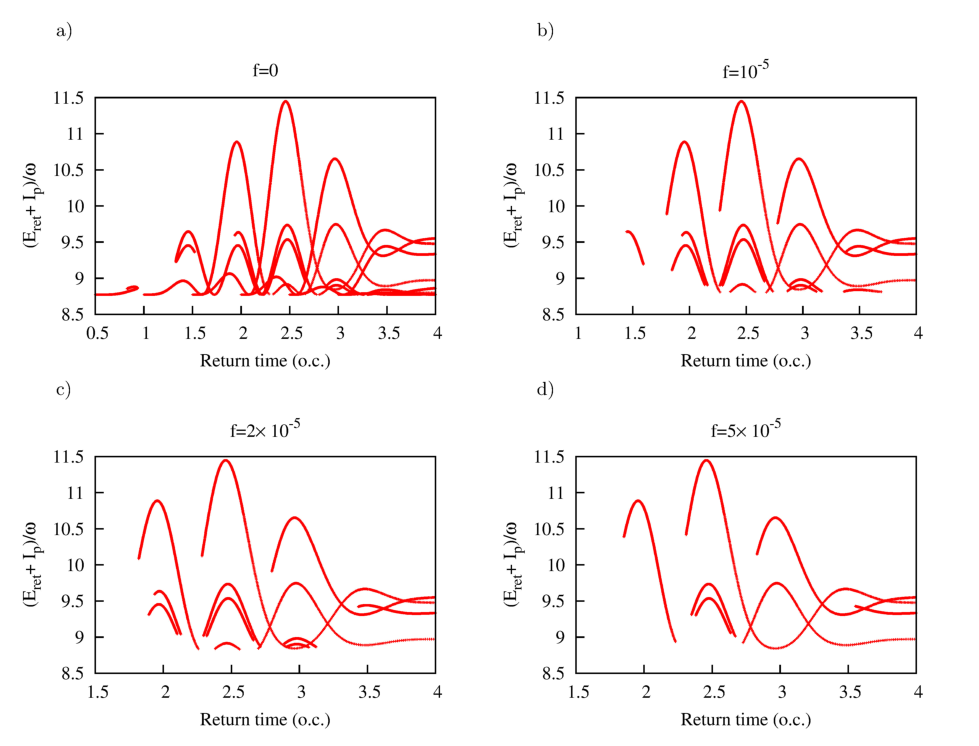}
\caption{(Color online) Classical calculations of 
emitted photon energy as function of return time for the dipole harmonic
radiation (a) and non-dipole harmonic radiation (b-d) with different filter
parameters.
}
\label{f7}
\end{figure*}

\begin{figure*}[h]%
\centering
\includegraphics[width=0.9\textwidth]{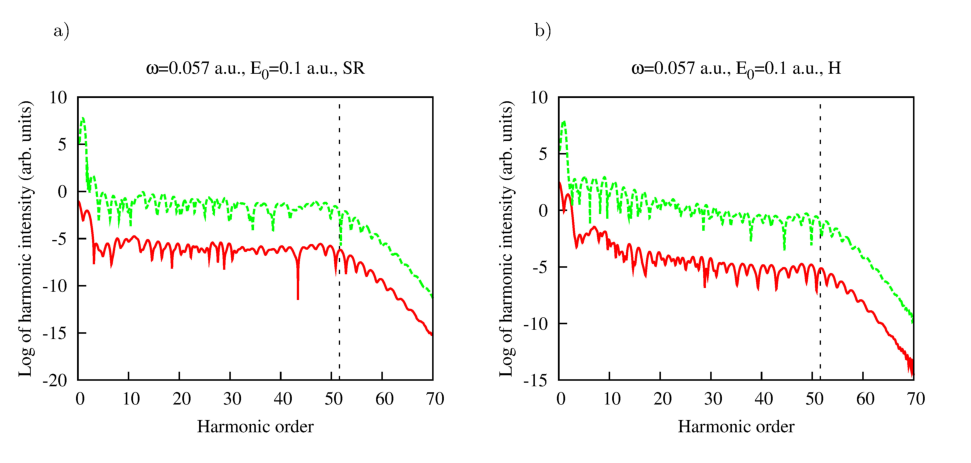}
\caption{(Color online) Dipole (dash green) and non-dipole (red solid) harmonic 
intensities for the pulse wavelength $\lambda=800$ nm. 
for the SR Yukawa and hydrogen atoms.
Vertical dash lines show cutoff positions.
}
\label{f8}
\end{figure*}

\begin{figure*}[h]%
\centering
\includegraphics[width=0.9\textwidth]{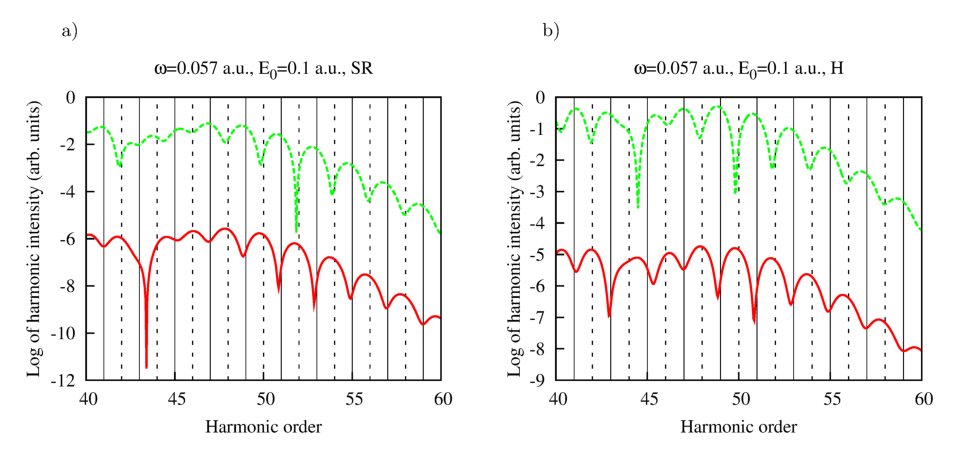}
\caption{(Color online) Dipole (dash green) and non-dipole (red solid) harmonic 
intensities for the pulse wavelength $\lambda=800$ nm for harmonics with orders 
$40 \le n \le 60$ for the SR Yukawa and hydrogen atoms.
Vertical solid and dash lines show positions of odd and even harmonics, respectively.
}
\label{f9}
\end{figure*}

\begin{figure*}[h]%
\centering
\includegraphics[width=0.9\textwidth]{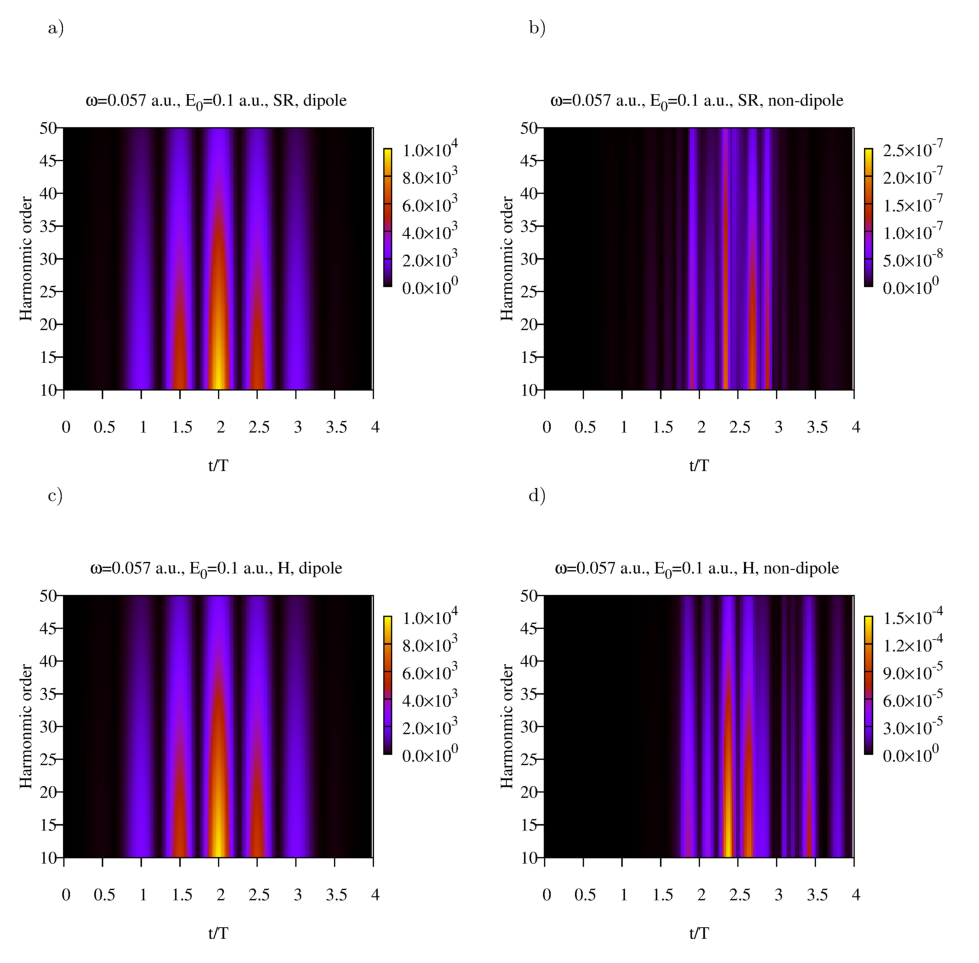}
\caption{(Color online) Gabor transform $|T(\Omega,t)|$ for 
the pulse wavelength $\lambda=800$ nm for
the SR Yukawa and hydrogen atoms.
}
\label{f10}
\end{figure*}

\begin{figure*}[h]
\bt{ll}
\hs{-1cm} a)   &
\hs{1cm} b)     \\
\hs{-2cm}\rs{80mm}{!}{\epsffile{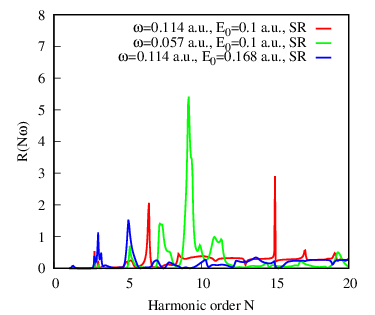}} &
\rs{80mm}{!}{\epsffile{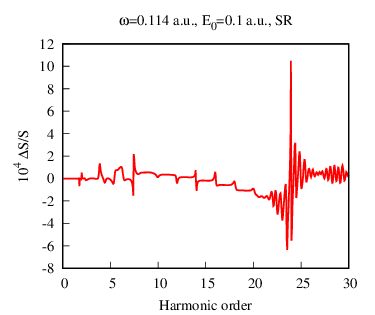}} \\
\end{tabular}
\caption{(Color online) (a) Estimate \eref{til2} for the ratio $R(\Omega)$.
(b) Normalized difference $\left(S_z(\Omega)-S^{\rm nr}_z(\Omega)\right)/S^{\rm nr}_z(\Omega)$
of the TDDE and TDSE calculations for the dipole harmonic intensities.}
\label{f31}
\end{figure*}

\begin{figure*}[h]
\bt{ll}
\hs{-1cm} a)   &
\hs{1cm} b)     \\
\hs{-2cm}\rs{80mm}{!}{\epsffile{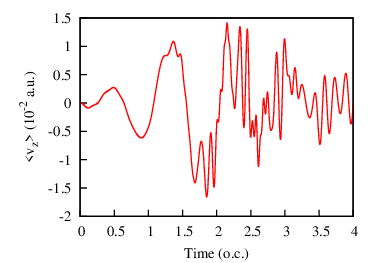}} &
\rs{80mm}{!}{\epsffile{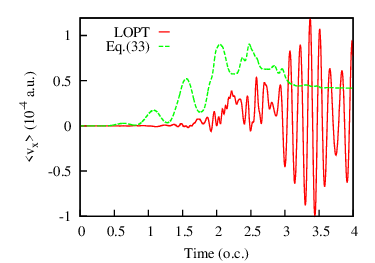}} \\
\end{tabular}
\caption{(Color online) (a) LOPT expectation value $v_z(t)$.
(b) $v_x(t)$ obtained in the LOPT calculation and using \Eref{vc}. 
Cosine pulse with $E_0=0.0534$ a.u., $\omega=0.057$ a.u.
has been used in the calculations.}
\label{f12}
\end{figure*}


\end{document}